\documentstyle[prl,aps]{revtex}
\begin{document}
\draft

\title{\bf RADIATION OF A QUANTUM LOCALIZED SOURCE}

\author{Alexander S. Shumovsky and Alexander A. Klyachko}

\address{Faculty of Science, Bilkent University, Bilkent, 06533
Ankara, Turkey}

\maketitle

\begin{abstract}
New effective  operators, describing the photons with given
polarization at given position with respect to a source are
proposed. These operators can be used to construct the near and
intermediate zones quantum optics. It is shown that the use of the
conventional plane photons can lead to a wrong results for quantum
fluctuations of polarization even in the far zone.
\end{abstract}


\narrowtext

\twocolumn


In the usual formulation of quantum optics, the radiation field is
considered in terms of states of photons with given energy and
linear momentum $\vec P$, corresponding to the translational
invariant solutions of the homogeneous wave equation (e.g., see
\cite{1,2}). This means that the quantum electromagnetic field is
described by the creation $a^+_{k \sigma}$ and annihilation $a_{k
\sigma}$ operators specified in the three dimensional space by
wave vector $\vec k$, determining direction of propagation, and by
two unit vectors of polarization ${\vec e}_{\sigma}$, $ \sigma
=1,2$, always orthogonal to $\vec k$. It is clear that this
representation is incapable of describing a radiation of a
localized quantum source - an atom, molecule, group of atoms, etc.
The point is that such a source emits a quantum multipole
radiation, corresponding to the rotational invariant solutions of
the wave equation. Therefore,  such a radiation must be described
in terms of states of photons with given energy and angular
momentum $\vec J$, in other words, in terms of the operators
$a^+_{jm}$ and $a_{jm}$, describing, at given $k$, creation and
annihilation of photons with given value $j$ of the angular
momentum and projection $m=-j, \cdots ,j$ \cite{3}. Since $[{\vec
J},{\vec P}]
\neq 0$, the two representations of the quantum electromagnetic field
are different in principle.

The use of plane waves of photons is usually argued by the fact
that  classical multipole radiation represented by the outgoing
spherical waves is well-approximated, at far distances from the source,
by  plane waves \cite{4}. Actually, it is not the case in the
quantum domain. This simplified picture overlooks the fact that the
plane and spherical (multipole) photons have different properties of
symmetry and therefore correspond to different physical quantities
which cannot be measured simultaneously. In particular, this difference is
manifested by the fact that the operators $a_{k \sigma}$ of
plane photons are always defined in the three-dimensional space,
while the operators $a_{jm}$ of spherical photons are determined in
the $(2j+1)$-dimensional space which coincides with the three-dimensional space
only in the case of dipole photons.

We now note that properties of the classical multipole radiation
strongly depend on position with respect to the source \cite{4}.
At the same time, the operators $a_{jm}$, describing the multipole
photons, are independent of position. The aim of this note is to
introduce a new representation of the spherical photon operators
which takes into account the spatial properties  as well as the
quantum nature  of the multipole radiation at any distance from
the source. As is already clear, these operators can be used to
extend the validity of the relations of conventional quantum
optics \cite{1,2} in the case of near and intermediate zone
radiation.

The spatial properties of a monochromatic pure multipole radiation of
a quantum source located at the origin can be described by the
following vector potential operator \cite{3}
\begin{eqnarray}
{\vec A}({\vec r})= \sum_{\mu =-1}^1 {\vec \chi}_{\mu}
\sum_{m=-j}^j V_{\mu m}({\vec r})a_{jm}
\label{1}
\end{eqnarray}
and by conjugated operator ${\vec A}^+({\vec r})$. Here $j$ is supposed to be
fixed and ${\vec \chi}_{\mu}$ is the complex unit vector, describing a spin
state of a photon with given projection $\mu =0, \pm 1$. The complex
coefficients $V_{\mu m}({\vec r})$ give the spatial dependence of
radiation. Their explicit form is well known \cite{3,4,5}. Since
\begin{eqnarray}
{\vec \chi}_{\mu}^+ \cdot {\vec \chi}_{\mu'} = \delta_{\mu \mu'},
\nonumber
\end{eqnarray}
we can introduce the spin components  of (1) as follows
\begin{eqnarray}
A_{ \mu}({\vec r})={\vec \chi}^+_{\mu} \cdot {\vec A}({\vec r}).
\label{2}
\end{eqnarray}
Since the quantum electrodynamics defines polarization as a
given spin state of photons \cite{5}, we can interpret (2) as the
component of the vector potential operator (1) with given
polarization. Here $\mu = \pm 1$ correspond to the circular
polarizations with opposite helicities, while $\mu =0$ specifies the
linear polarization in the radial direction which always exists in
the multipole radiation \cite{4}.

Let us stress that we choose here not the field {\it per se}, but
vector potential (1) because of the following reasons.  First,
${\vec A}({\vec r})$ obey the gauge invariance. Second, the
polarization is usually determined to be the measure of
transversal anisotropy of electric field or magnetic induction,
depending on the type of radiation either magnetic or electric
\cite{1,4}. Since the information about this anisotropy is
described by $V_{\mu m}({\vec r})$ in (1), the use of vector
potential permits us to consider both types of the multipole
radiation simultaneously. Taking into account that
\begin{eqnarray}
[a_{jm},a^+_{jm'}]= \delta_{mm'} , \nonumber
\end{eqnarray}
we get
\begin{eqnarray}
[A_{ \mu}({\vec r}),A^+_{ \mu'}({\vec r})]= \sum_{m=-j}^j V_{\mu
m}({\vec r})V^*_{\mu' m}({\vec r}) \equiv {\cal V}_{\mu \mu'}({\vec
r}).
\label{3}
\end{eqnarray}
At first sight, the right-hand side of (3) determines the entries
of a nonnegative  $(3 \times 3)$ Hermitian matrix at any point
$\vec r$. Moreover $\cite{6}$ the matrix $\cal V=||\cal
V_{\mu\mu^\prime}||$  is real symmetric  for a proper choice of
the longitude angle $\varphi$, which is irrelevant,  and thus may
be diagonalized by a suitable rotation $U\in \mbox{SO(3)}$. The
same rotation converts the operator (2) into the form
\begin{eqnarray}
a_{\mu}({\vec r})=(W_{\mu}({\vec r}))^{-1/2} \sum_{\mu' =-1}^1
U^*_{\mu \mu'}({\vec r})A_{\mu'}({\vec r}),
\label{4}
\end{eqnarray}
where $W_{\mu}({\vec r})\ge0$  denotes eigenvalues of $\cal
V=||\cal V_{\mu\mu^\prime}||$. It can be easily seen that
\begin{eqnarray}
[a_{\mu}({\vec r}),a^+_{\mu'}({\vec r})]= \delta_{\mu \mu'},
\label{5}
\end{eqnarray}
so that the operators (4) are similar to the standard photon
operators. In other words, the operators (4) form a representation
of the Weyl-Heisenberg algebra of photons at any point $\vec r$ of
the three dimensional space. By virtue of the above discussion, we
can choose to interpret $a^+_{\mu}({\vec r})$ and $a_{\mu}({\vec
r})$ as the effective creation and annihilation operators of a
photon with polarization $\mu$ at the point $\vec r$. The
operators (4) are linear forms in $a_{jm}$. This dependence
reflects the fact that the orbital part of the angular momentum of
photons cannot be well defined in the states with given spin
(polarization) \cite{5}.

We now note that the matrix $\cal V$ in (3) has very simple physical meaning.
In fact, the form
\begin{eqnarray}
P_{\mu' \mu}^{(n)}({\vec r}) \equiv \langle A^+_{\mu'}({\vec r})A_{\mu}({\vec r}) \rangle
\nonumber
\end{eqnarray}
can be considered as the quantum counterpart of the polarization matrix for the
multipole radiation at an arbitrary point \cite{7}, corresponding to the
normal order of operators. In turn,
\begin{eqnarray}
P^{an}_{\mu' \mu}({\vec r}) \equiv \langle A_{\mu}({\vec
r})A^+_{\mu'}({\vec r}) \rangle
\nonumber
\end{eqnarray}
represents the quantum polarization matrix, determined in terms of
an antinormal order. Since the order of factors is inessential in
the classical definition of the polarization matrix \cite{1}, the
matrix
\begin{eqnarray}
{\cal V}_{\mu \mu'}({\vec r}) = P^{(an)}_{\mu' \mu}({\vec r})-P^{(n)}_{\mu' \mu}({\vec r}) \nonumber
\end{eqnarray}
in (3) describes the deviation of the quantum polarization matrix from the
classical one. In other words, the matrix ${\cal V}({\vec r})$ describes the
quantum fluctuations of the elements of the polarization matrix at any point.

The physical picture of the multipole radiation of a quantum
source such as an atom, can be now described as follows. In the
generation zone with $r=r_a \sim 10^{-8}cm$, the process or
radiation is described in terms of the operators $a^+_{jm},a_{jm}$
within the framework of some model such as the Jaynes-Cummings
Hamiltonian (e.g., see \cite{8}) for a multipole transition
\cite{9}. Everywhere outside the generation zone, the radiation is described
by the effective operators (4).

The use of the operators (4) enables us to extend the known
notions of quantum optics in the case of radiation of a quantum
localized source at any distance. For example, the type of
statistics of photons \cite{1} with given polarization at given
point can be described by the following ``local" Mandel's
$Q$-parameter
\begin{eqnarray}
Q_{\mu}({\vec r})= \frac{\langle[ \Delta n_{\mu}({\vec r})]^2
\rangle - \langle n_{\mu}({\vec r}) \rangle}{\langle n_{\mu}({\vec
r}) \rangle} ,
\label{6}
\end{eqnarray}
where the operator
\begin{eqnarray}
n_{\mu}({\vec r})=a^+_{\mu}({\vec r})a_{\mu}({\vec r}) \nonumber
\end{eqnarray}
describes the number of photons with given polarization at the point
$\vec r$. In turn, the coherent states with given polarization can be defined
with the aid of (4) as follows
\begin{eqnarray}
a_{\mu}({\vec r})| \alpha  \rangle = \alpha_{\mu} ({\vec
r}) | \alpha \rangle ,
\label{7}
\end{eqnarray}
so that
\begin{eqnarray}
| \alpha  \rangle = \bigotimes_{m=-j}^j | \alpha_m \rangle ,
\nonumber
\end{eqnarray}
and
\begin{eqnarray}
\alpha_{\mu}({\vec r})=[W_{\mu}({\vec r})]^{-1/2} \sum_{\mu' =-1}^1
\sum_{m=-j}^j U^*_{\mu \mu'}({\vec r})V_{\mu' m}({\vec r}) \alpha_m ,
\nonumber
\end{eqnarray}
where $| \alpha_m \rangle$ is the standard coherent state defined
by the relation
\begin{eqnarray}
a_{jm}| \alpha_m \rangle =  \alpha_m | \alpha_m \rangle .
\nonumber
\end{eqnarray}
It is clear that $Q_{\mu}({\vec r})=0$ everywhere in the coherent
state (7). This means that coherence is a global property of the
radiation field although the parameter of coherence in (7) changes
with the distance. To trace this change, consider an electric dipole
radiation with $j=1$ and $m=0, \pm 1$ in the coherent state (7).
Assume that $\alpha_m = \alpha \delta_{m,+1}$. In view of the
asymptotics of $V_{\mu m}({\vec r})$ in (1) \cite{3,4,5}, for the
position dependent parameter of the coherent state (7) we get
\begin{eqnarray}
\alpha_{\mu}(0)= \alpha \delta_{\mu, +1} \nonumber
\end{eqnarray}
so that the source, localized at the origin, generates the
circularly polarized coherent light with positive helicity. It
follows from (7) that the radiation in the near and intermediate
zones consists of all three components in the coherent state each.
Since $V_{0,m}({\vec r})$ vanishes at far distance \cite{3,4}, the
far-zone radiation consists of only two coherent transversal
components such that
\begin{eqnarray}
\frac{\alpha_{-1} ({\vec r})}{\alpha_{+1} ({\vec r})} \rightarrow
\frac{1- \cos^2 \theta}{1+ \cos^2 \theta} e^{2i \phi}, \nonumber
\end{eqnarray}
where $\theta$ and $\phi$ are the vectorial angles in the spherical
coordinates of the point $\vec r$. Other basic relations of the
quantum optics can be modified in a similar way. Hence, the operators
(4) can be used to construct the near and intermediate field quantum
optics. Let us stress that  conventional near field optics deals with
the classical field (e.g., see \cite{10}).

Consider now the polarization properties of the quantum multipole
radiation. It is customary to discuss  the polarization in terms
of the polarization matrix \cite{1} and corresponding Stokes parameters
(Stokes operators, in the quantum domain \cite{11}) . In the case of a
multipole radiation, this is the $(3 \times 3)$ Hermitian matrix
\cite{9}.  The generalized Stokes operators \cite{9} can be
constructed either from the quantum counterpart of the elements of
this matrix \cite{9,12} or from the generators of the $SU(3)$
sub-algebra in the Weyl-Heisenberg algebra of photons \cite{13},
describing the independent Hermitian bilinear forms in the creation
and annihilation operators. We restrict our consideration by only two
generalized Stokes operators
\begin{eqnarray}
S_1({\vec r}) & = & a^+_+a_0+a^+_0a_-+a^+_-a_+ +H.c., \nonumber \\
S_2({\vec r}) & = & -i[a^+_+a_0+a^+_0a_-+a^+_-a_+ -H.c.].
\label{8}
\end{eqnarray}
describing the phase information \cite{9,12}. Here $a_{\pm ,0}$ denotes the
operators (4) at $\mu = \pm 1,0$. It is clear that
$[S_1({\vec r}),S_2({\vec r})]=0$, so that corresponding
physical quantities can be measured at once at any point. Since
$V_{0m}({\vec r})$ vanishes at far distance \cite{4}, the intensity
of linearly polarized radial component of the multipole radiation
tends to zero as $r \rightarrow 0$. This means that at far distances
the field with $\mu =0$ should be considered in the vacuum state.
Omitting for the moment the position dependence in (4), for the
averages of (8) over the far-zone radiation we get
\begin{eqnarray}
\langle S_1 \rangle & = & 2Re( \langle a^+_-a_+ \rangle ), \nonumber
\\
\langle S_2 \rangle & = & 2Im( \langle a^+_-a_+ \rangle ), \nonumber
\end{eqnarray}
which coincides with the standard phase-dependent Stokes parameters
determined in the circular polarization basis \cite{1,11}. These two
parameters, apart from an intensity-dependent factor, give the
cosine and sine of the phase difference between two circularly
polarized components. Hence, the polarization of quantum multipole
radiation at far distances looks like that of the plane waves of
photons.

At the same time, the fluctuations of (8) at far distance are
different from those described by the conventional Stokes operators
\cite{9}. Taking into account the commutation relations (5), for the
variance of $S_1({\vec r})$ in (8) at far distance we get
\begin{eqnarray}
\langle ( \Delta S_1)^2 \rangle = 2Re \langle ( \Delta a^+_-a_+)^2
\rangle +2( \langle n_+n_- \rangle -| \langle a^+_+a_- \rangle |^2) \nonumber \\+
\langle n_+ \rangle
+ \langle n_- \rangle
+2Re \langle a^+_+a_- \rangle + \langle n_+ \rangle + \langle n_-
\rangle .
\label{9}
\end{eqnarray}
Here the first four terms coincide with the standard result
calculated as though the radiation is represented by the plane
photons everywhere. The additional three terms come from the
commutation relations (5). They reflects the rotational invariance
of the states of multipole photons. The presence of these terms
increases the quantum fluctuations of transversal polarization and
changes them qualitatively because the term $Re \langle a^+_+a_-
\rangle$ involves an additional phase dependence. The above result
shows that the use of an approximation by the plane waves of
photons rather than the spherical photons can lead to a wrong
results even in the far zone. Let us stress here that the correct
estimation of quantum fluctuations of polarization is very
important, especially for the quantum entanglement investigation.

In summary, we report here the definition of the effective photon operators (4),
describing the photons with given polarization at any distance from emitted
a quantum localized source located at the origin. These operators
can be used in order to extend the validity of relations of conventional quantum optics in the near zone and intermediate zone
 as well as to give the correct description of  quantum
fluctuation of physical quantities at far distances.

\end{document}